\pdfoutput=1
\documentclass[12pt,a4paper]{article}

\usepackage{graphicx}
\usepackage{xspace} % to avoid problems with missing or double spaces after predefined symbols
\usepackage{color}
\usepackage{colortbl}
\usepackage{amsmath}
\usepackage{ifthen}
\usepackage{multirow}
\usepackage{caption}
\usepackage{dcolumn}
\usepackage{booktabs}
\usepackage{amssymb}
\usepackage{amsfonts}
\usepackage{upgreek} % Adds in support for greek letters in roman typeset
\usepackage{hyperref} % Hyperlinks in references
\usepackage[all]{hypcap} % Internal hyperlinks to floats.
\usepackage{cite} % Allows for ranges in citations
\usepackage{mciteplus}

\newboolean{pdflatex}
\setboolean{pdflatex}{true} 

\newboolean{articletitles}
\setboolean{articletitles}{true} % False removes titles in references

\newboolean{uprightparticles}
\setboolean{uprightparticles}{false} % Set to true to get roman particle symbols

\def\lhcb {\mbox{LHCb}\xspace}
\def\ux85 {\mbox{UX85}\xspace}

\ifthenelse{\boolean{uprightparticles}}%
{

 \def\Pmu         {\ensuremath{\upmu}\xspace}

 \def\Ppi         {\ensuremath{\uppi}\xspace}

 \def\Ppsi        {\ensuremath{\uppsi}\xspace}

 \def\PDelta      {\ensuremath{\Delta}\xspace}                 
 \def\PXi      {\ensuremath{\Xi}\xspace}                 
 \def\PLambda      {\ensuremath{\Lambda}\xspace}                 
 \def\PSigma      {\ensuremath{\Sigma}\xspace}                 
 \def\POmega      {\ensuremath{\Omega}\xspace}                 
 \def\PUpsilon      {\ensuremath{\Upsilon}\xspace}                 
 
 \def\PB      {\ensuremath{\mathrm{B}}\xspace}                 
                  
 \def\PD      {\ensuremath{\mathrm{D}}\xspace}

 \def\PJ      {\ensuremath{\mathrm{J}}\xspace}                 
 \def\PK      {\ensuremath{\mathrm{K}}\xspace}

 \def\Pb      {\ensuremath{\mathrm{b}}\xspace}                 
 \def\Pc      {\ensuremath{\mathrm{c}}\xspace}

 \def\Pi      {\ensuremath{\mathrm{i}}\xspace}

 \def\Pp      {\ensuremath{\mathrm{p}}\xspace}

 \def\Ps      {\ensuremath{\mathrm{s}}\xspace}

}
{

 \def\Pmu         {\ensuremath{\mu}\xspace}

 \def\Ppi         {\ensuremath{\pi}\xspace}

 \def\Ppsi        {\ensuremath{\psi}\xspace}                 
                  
 \mathchardef\PDelta="7101
 \mathchardef\PXi="7104
 \mathchardef\PLambda="7103
 \mathchardef\PSigma="7106
 \mathchardef\POmega="710A
 \mathchardef\PUpsilon="7107
                  
 \def\PB      {\ensuremath{B}\xspace}                 
                  
 \def\PD      {\ensuremath{D}\xspace}

 \def\PJ      {\ensuremath{J}\xspace}                 
 \def\PK      {\ensuremath{K}\xspace}

 \def\Pb      {\ensuremath{b}\xspace}                 
 \def\Pc      {\ensuremath{c}\xspace}

 \def\Pi      {\ensuremath{i}\xspace}

 \def\Pp      {\ensuremath{p}\xspace}

 \def\Ps      {\ensuremath{s}\xspace}

}

\def\mup        {\ensuremath{\Pmu^+}\xspace}
\def\mun        {\ensuremath{\Pmu^-}\xspace} % muon negative (\mum is taken)
\def\mumu       {\ensuremath{\Pmu^+\Pmu^-}\xspace}

\def\squark    {\ensuremath{\Ps}\xspace}

\def\cquark    {\ensuremath{\Pc}\xspace}

\def\bquark    {\ensuremath{\Pb}\xspace}

\def\pion  {\ensuremath{\Ppi}\xspace}

\def\pim   {\ensuremath{\pion^-}\xspace}
\def\pipi  {\ensuremath{\pion^+\pion^-}\xspace}

\def\pimp  {\ensuremath{\pion^\mp}\xspace}

\def\kaon  {\ensuremath{\PK}\xspace}
  \def\Kbar  {\kern 0.2em\overline{\kern -0.2em \PK}{}\xspace}

\def\Kz    {\ensuremath{\kaon^0}\xspace}
\def\Kzb   {\ensuremath{\Kbar^0}\xspace}
\def\KzKzb {\ensuremath{\Kz \kern -0.16em \Kzb}\xspace}
\def\Kp    {\ensuremath{\kaon^+}\xspace}
\def\Km    {\ensuremath{\kaon^-}\xspace}
\def\Kpm   {\ensuremath{\kaon^\pm}\xspace}

\def\KpKm  {\ensuremath{\Kp \kern -0.16em \Km}\xspace}
\def\KS    {\ensuremath{\kaon^0_{\rm\scriptscriptstyle S}}\xspace} 
 
\def\Kstarz  {\ensuremath{\kaon^{*0}}\xspace}

  \def\Dbar    {\kern 0.2em\overline{\kern -0.2em \PD}{}\xspace}
\def\D       {\ensuremath{\PD}\xspace}

\def\Dz      {\ensuremath{\D^0}\xspace}
\def\Dzb     {\ensuremath{\Dbar^0}\xspace}
\def\DzDzb   {\ensuremath{\Dz {\kern -0.16em \Dzb}}\xspace}
\def\Dp      {\ensuremath{\D^+}\xspace}
\def\Dm      {\ensuremath{\D^-}\xspace}

\def\DpDm    {\ensuremath{\Dp {\kern -0.16em \Dm}}\xspace}

\def\B       {\ensuremath{\PB}\xspace}
\def\Bbar    {\ensuremath{\kern 0.18em\overline{\kern -0.18em \PB}{}}\xspace}

\def\Bu      {\ensuremath{\B^+}\xspace}
\def\Bub     {\ensuremath{\B^-}\xspace}

\def\Bd      {\ensuremath{\B^0}\xspace}
\def\Bs      {\ensuremath{\B^0_\squark}\xspace}
\def\Bsb     {\ensuremath{\Bbar^0_\squark}\xspace}
\def\Bdb     {\ensuremath{\Bbar^0}\xspace}

\def\jpsi     {\ensuremath{{\PJ\mskip -3mu/\mskip -2mu\Ppsi\mskip 2mu}}\xspace}

  \def\Y#1S{\ensuremath{\PUpsilon{(#1S)}}\xspace}% no space before {...}!

\def\FourS {\Y4S}

\def\proton      {\ensuremath{\Pp}\xspace}

\def\L {\ensuremath{\PLambda}\xspace}
\def\Lbar {\ensuremath{\kern 0.1em\overline{\kern -0.1em\PLambda}}\xspace}

\def\BF         {{\ensuremath{\cal B}\xspace}}

\def\BR         {\BF}
\def\to                 {\ensuremath{\rightarrow}\xspace}

\newcommand{\tauBs}{\ensuremath{\tau_{\Bs}}\xspace}

\def\CP                {\ensuremath{C\!P}\xspace}

\def\AT#1     {\ensuremath{A_{\mathrm{T}}^{#1}}\xspace}           % 2

\def\C#1      {\ensuremath{\mathcal{C}_{#1}}\xspace}                       % 9
\def\Cp#1     {\ensuremath{\mathcal{C}_{#1}^{'}}\xspace}                    % 7
\def\Ceff#1   {\ensuremath{\mathcal{C}_{#1}^{\mathrm{(eff)}}}\xspace}        % 9  
\def\Cpeff#1  {\ensuremath{\mathcal{C}_{#1}^{'\mathrm{(eff)}}}\xspace}       % 7
\def\Ope#1    {\ensuremath{\mathcal{O}_{#1}}\xspace}                       % 2
\def\Opep#1   {\ensuremath{\mathcal{O}_{#1}^{'}}\xspace}                    % 7

\newcommand{\tev}{\ensuremath{\mathrm{\,Te\kern -0.1em V}}\xspace}
\newcommand{\gev}{\ensuremath{\mathrm{\,Ge\kern -0.1em V}}\xspace}
\newcommand{\mev}{\ensuremath{\mathrm{\,Me\kern -0.1em V}}\xspace}
\newcommand{\kev}{\ensuremath{\mathrm{\,ke\kern -0.1em V}}\xspace}
\newcommand{\ev}{\ensuremath{\mathrm{\,e\kern -0.1em V}}\xspace}
\newcommand{\gevc}{\ensuremath{{\mathrm{\,Ge\kern -0.1em V\!/}c}}\xspace}
\newcommand{\mevc}{\ensuremath{{\mathrm{\,Me\kern -0.1em V\!/}c}}\xspace}
\newcommand{\gevcc}{\ensuremath{{\mathrm{\,Ge\kern -0.1em V\!/}c^2}}\xspace}
\newcommand{\gevgevcccc}{\ensuremath{{\mathrm{\,Ge\kern -0.1em V^2\!/}c^4}}\xspace}
\newcommand{\mevcc}{\ensuremath{{\mathrm{\,Me\kern -0.1em V\!/}c^2}}\xspace}

\def\mum  {\ensuremath{\,\upmu\rm m}\xspace}

\def\invfb   {\ensuremath{\mbox{\,fb}^{-1}}\xspace}

\def\ps   {\ensuremath{{\rm \,ps}}\xspace}

\def\gsim{{~\raise.15em\hbox{$>$}\kern-.85em
          \lower.35em\hbox{$\sim$}~}\xspace}
\def\lsim{{~\raise.15em\hbox{$<$}\kern-.85em
          \lower.35em\hbox{$\sim$}~}\xspace}

\def\sPlot{\mbox{\em sPlot}}

\def\pt         {\mbox{$p_{\rm T}$}\xspace}

\def\evtgen     {\mbox{\textsc{EvtGen}}\xspace}
\def\pythia     {\mbox{\textsc{Pythia}}\xspace}

\def\geant      {\mbox{\textsc{Geant4}}\xspace}

\def\photos     {\mbox{\textsc{Photos}}\xspace}

\def\tell1  {TELL1\xspace}
\def\ukl1   {UKL1\xspace}

\newcolumntype{P}[1]{D{!}{\ \pm\ }{#1}}

\renewcommand{\sPlot}{\emph{sPlot}\xspace}

\def \U4S{\ensuremath{1.055 \pm 0.025}\xspace}
\def \BFBd{8.98}
\def \BFBdE{0.35}

\def \tauBs{\ensuremath{1.513 \pm 0.011\:\text{ps}}\xspace}
\def \DeltaGamma{\ensuremath{0.106 \pm 0.013\:\text{ps}^{-1}}\xspace}
\def \ADGSM{0.944}
\def \ADGSME{0.066}
\def \tauEffSM{\ensuremath{1.639 \pm 0.022\:\text{ps}}\xspace}
\def \fsfd{\ensuremath{0.256 \pm 0.020}\xspace}

\def \SystAcc{1.000}
\def \SystAccE{0.005}
\def \SystBkg{1.000}
\def \SystBkgE{0.039}
\def \SystMass{1.000}
\def \SystMassE{0.004}
\def \SystBdLife{1.000}
\def \SystBdLifeE{0.005}
\def \SystFitMethod{1.002}
\def \SystFitMethodE{0.002}
\def \SystTauDef{0.999}
\def \SystTauDefE{0.001}
\def \SystTauTot{1.001}
\def \SystTauTotE{0.040}

\def \EffRaw{1.75}
\def \EffRawE{0.12}
\def \TauBsEff{1.75}
\def \TauBsEffStat{0.12}
\def \TauBsEffSyst{0.07}

\def \YRatio{0.0116}
\def \YRatioE{0.0008}

\def \SystFitModel{1.000}
\def \SystFitModelE{0.034}
\def \SystMassRes{1.000}
\def \SystMassResE{0.014}
\def \SystSelEffCorr{0.968}
\def \SystSelEffCorrE{0.007}
\def \SystTot{0.968}
\def \SystTotE{0.034}

\def \BRBsBd{0.0439}
\def \BRBsBdStat{0.0032}
\def \BRBsBdSyst{0.0015}
\def \BRBsBdfds{0.0034}

\def \BFBs{1.97}
\def \BFBsStat{0.14}
\def \BFBsSyst{0.07}
\def \BFBsfds{0.15}
\def \BFBsPDG{0.08}
\def \BFBsTot{0.23}
\begin{document}

%%%%%%%%%%%%%%%%%%%%%%%%%%%%%%%%%%%%%%%%%%%%%%%%%%%%%%%%%%%%%%%%%%%%%%%%%%%
%%%%% TITLE %%%%%
%%%%%%%%%%%%%%%%%%%%%%%%%%%%%%%%%%%%%%%%%%%%%%%%%%%%%%%%%%%%%%%%%%%%%%%%%%%
\renewcommand{\thefootnote}{\fnsymbol{footnote}}
\setcounter{footnote}{1}
%%%%%%%%%%%%%%%%%%%%%%%%%%%%%%%%%%%%%%%%%%%%%%%%%%%%%%%%%%%%%%%%%%%%%%%%%%%
%%%%%  TITLE PAGE  %%%%%%
%%%%%%%%%%%%%%%%%%%%%%%%%%%%%%%%%%%%%%%%%%%%%%%%%%%%%%%%%%%%%%%%%%%%%%%%%%%
\begin{titlepage}
\pagenumbering{roman}

% =====
% Header
\vspace*{-1.5cm}
\centerline{\large EUROPEAN ORGANIZATION FOR NUCLEAR RESEARCH (CERN)}
\vspace*{1.5cm}
\hspace*{-0.5cm}
\begin{tabular*}{\linewidth}{lc@{\extracolsep{\fill}}r}
\vspace*{-2.7cm}\mbox{\!\!\!\includegraphics[width=.14\textwidth]{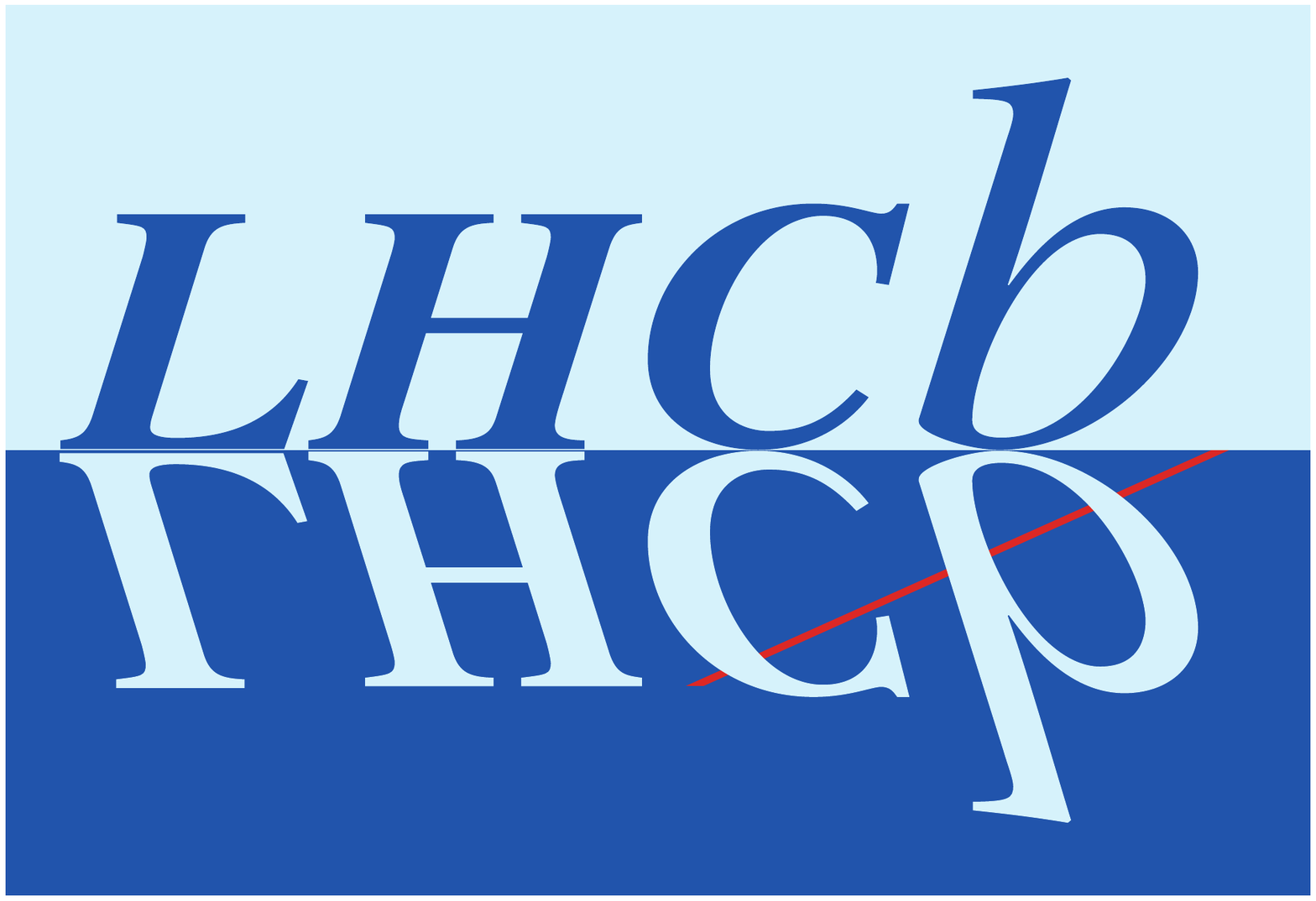}} & & \\ % Logo
 & & CERN-PH-EP-2013-063 \\  % ID 
 & & LHCb-PAPER-2013-015 \\  % ID 
 & & April 16, 2013 % Date
\end{tabular*}

\vspace*{4.0cm}

% =====
% Title
{\bf\boldmath\huge
\begin{center}
Measurement of the effective \mbox{\Bs\to\jpsi{}\KS} lifetime
\end{center}
}

\vspace*{2.0cm}

% =====
% Authors
\begin{center}
The LHCb collaboration\footnote{Authors are listed on the following pages.}
\end{center}

\vspace{\fill}

% =====
% Abstract
\begin{abstract}
\noindent
This paper reports the first measurement of the effective \Bs\to\jpsi{}\KS lifetime
and an updated measurement of its time-integrated branching fraction.
Both measurements are performed with a data sample, corresponding to an integrated luminosity of 1.0\:\invfb of
$pp$ collisions, recorded by the LHCb experiment in 2011 at a centre-of-mass energy of $7\:\tev$.
The results are:
$\tau_{\jpsi{}\KS}^{\text{eff}} = \TauBsEff \pm \TauBsEffStat\:(\text{stat}) \pm \TauBsEffSyst\:(\text{syst})\:\text{ps}$
and
\mbox{$\BF(\Bs\to\jpsi{}\KS)=(\BFBs\pm \BFBsTot)\times10^{-5}$}.
For the latter measurement, the uncertainty includes both statistical and systematic sources.
\end{abstract}

\vspace*{2.0cm}

\begin{center}
Published in Nucl.\ Phys.\ B
\end{center}

\vspace{\fill}

{\footnotesize 
\centerline{\copyright~CERN on behalf of the \lhcb collaboration, license \href{http://creativecommons.org/licenses/by/3.0/}{CC-BY-3.0}.}}
\vspace*{2mm}

\end{titlepage}

%%%%%%%%%%%%%%%%%%%%%%%%%%%%%%%%%%%%%%%%%%%%%%%%%%%%%%%%%%%%%%%%%%%%%%%%%%%
%%%%%  EOD OF TITLE PAGE  %%%%%%
%%%%%%%%%%%%%%%%%%%%%%%%%%%%%%%%%%%%%%%%%%%%%%%%%%%%%%%%%%%%%%%%%%%%%%%%%%%

% =====
%  empty page follows the title page
\newpage
\setcounter{page}{2}
\mbox{~}
\newpage

% =====
% Author List
%%%%%%%%%%%%%%%%%%%%%%%%%%%%%%%%%%%%%%%%%%
\centerline{\large\bf LHCb collaboration}
\begin{flushleft}
\small
R.~Aaij$^{40}$, 
C.~Abellan~Beteta$^{35,n}$, 
B.~Adeva$^{36}$, 
M.~Adinolfi$^{45}$, 
C.~Adrover$^{6}$, 
A.~Affolder$^{51}$, 
Z.~Ajaltouni$^{5}$, 
J.~Albrecht$^{9}$, 
F.~Alessio$^{37}$, 
M.~Alexander$^{50}$
% ...

{\footnotesize \it
% ...
$ ^{5}$Clermont Universit\'{e}, Universit\'{e} Blaise Pascal, CNRS/IN2P3, LPC, Clermont-Ferrand, France\\
$ ^{6}$CPPM, Aix-Marseille Universit\'{e}, CNRS/IN2P3, Marseille, France\\
% ...
$ ^{9}$Fakult\"{a}t Physik, Technische Universit\"{a}t Dortmund, Dortmund, Germany\\
% ...
$ ^{35}$Universitat de Barcelona, Barcelona, Spain\\
$ ^{36}$Universidad de Santiago de Compostela, Santiago de Compostela, Spain\\
$ ^{37}$European Organization for Nuclear Research (CERN), Geneva, Switzerland\\
% ...
$ ^{40}$Nikhef National Institute for Subatomic Physics, Amsterdam, The Netherlands\\
% ...
$ ^{45}$H.H. Wills Physics Laboratory, University of Bristol, Bristol, United Kingdom\\
% ...
$ ^{50}$School of Physics and Astronomy, University of Glasgow, Glasgow, United Kingdom\\
$ ^{51}$Oliver Lodge Laboratory, University of Liverpool, Liverpool, United Kingdom\\
% ...
$ ^{n}$LIFAELS, La Salle, Universitat Ramon Llull, Barcelona, Spain\\
% ...
}
\end{flushleft}
%%%%%%%%%%%%%%%%%%%%%%%%%%%%%%%%%%%%%%%%%%

\cleardoublepage

%%%%%%%%%%%%%%%%%%%%%%%%%%%%%%%%%%%%%%%%%%%%%%%%%%%%%%%%%%%%%%%%%%%%%%%%%%%
%%%%% MAIN TEXT %%%%%
%%%%%%%%%%%%%%%%%%%%%%%%%%%%%%%%%%%%%%%%%%%%%%%%%%%%%%%%%%%%%%%%%%%%%%%%%%%

\pagestyle{plain} % restore page numbers for the main text
\setcounter{page}{1}
\pagenumbering{arabic}

% Introduction
%%%%%%%%%%%%%%%%%%%%%%%%%%%%%%%%%%%%%%%%%%%%%%%%%%%%%%%%%%%%%%%%%%%%%%%%%%%
\section{Introduction}\label{Sec:Introduction}
%%%%%%%%%%%%%%%%%%%%%%%%%%%%%%%%%%%%%%%%%%%%%%%%%%%%%%%%%%%%%%%%%%%%%%%%%%%

% =====
% Intro from Bs2JpsiPhi paper
In the Standard Model (SM), \CP{} violation arises through a single
phase in the CKM quark mixing matrix~\cite{Kobayashi:1973fv,*Cabibbo:1963yz}.  
In decays of neutral \B mesons (\B stands for a \Bd or \Bs meson)
to a final state accessible to both $\B$ and $\Bbar$, 
the interference between the amplitude for the direct decay
and the amplitude for decay via oscillation leads to
time-dependent \CP{} violation. 
% =====
% Why Bs2JpsiKs
A measurement of the time-dependent \CP{} asymmetry in the \Bd\to\jpsi{}\KS mode allows for a determination
of the $\Bd$--$\Bdb$ mixing phase $\phi_d$.
In the SM it is equal to $2\beta$~\cite{Bigi:1981qs}, where $\beta$ is one of the angles of the unitarity triangle
in the quark mixing matrix.
This phase has already been well measured by the $B$ factories \cite{Adachi:2012et, Babar:2009yr},
but further improvements are still necessary to conclusively resolve
possible small tensions with the other measurements constraining the unitarity triangle
\cite{Bona:2005vz, *Charles:2004jd}.
The latest average composed by the Heavy Flavour Averaging Group (HFAG) is
$\sin\phi_d = 0.682 \pm 0.019$ \cite{HFAG}.
To achieve precision below the percent level, knowledge of the doubly Cabibbo-suppressed higher order	perturbative corrections, originating from penguin topologies, becomes mandatory.
These contributions are difficult to calculate reliably and therefore need
to be determined directly from experimentally accessible observables.

\begin{figure}[t]
\includegraphics[width=0.49\textwidth]{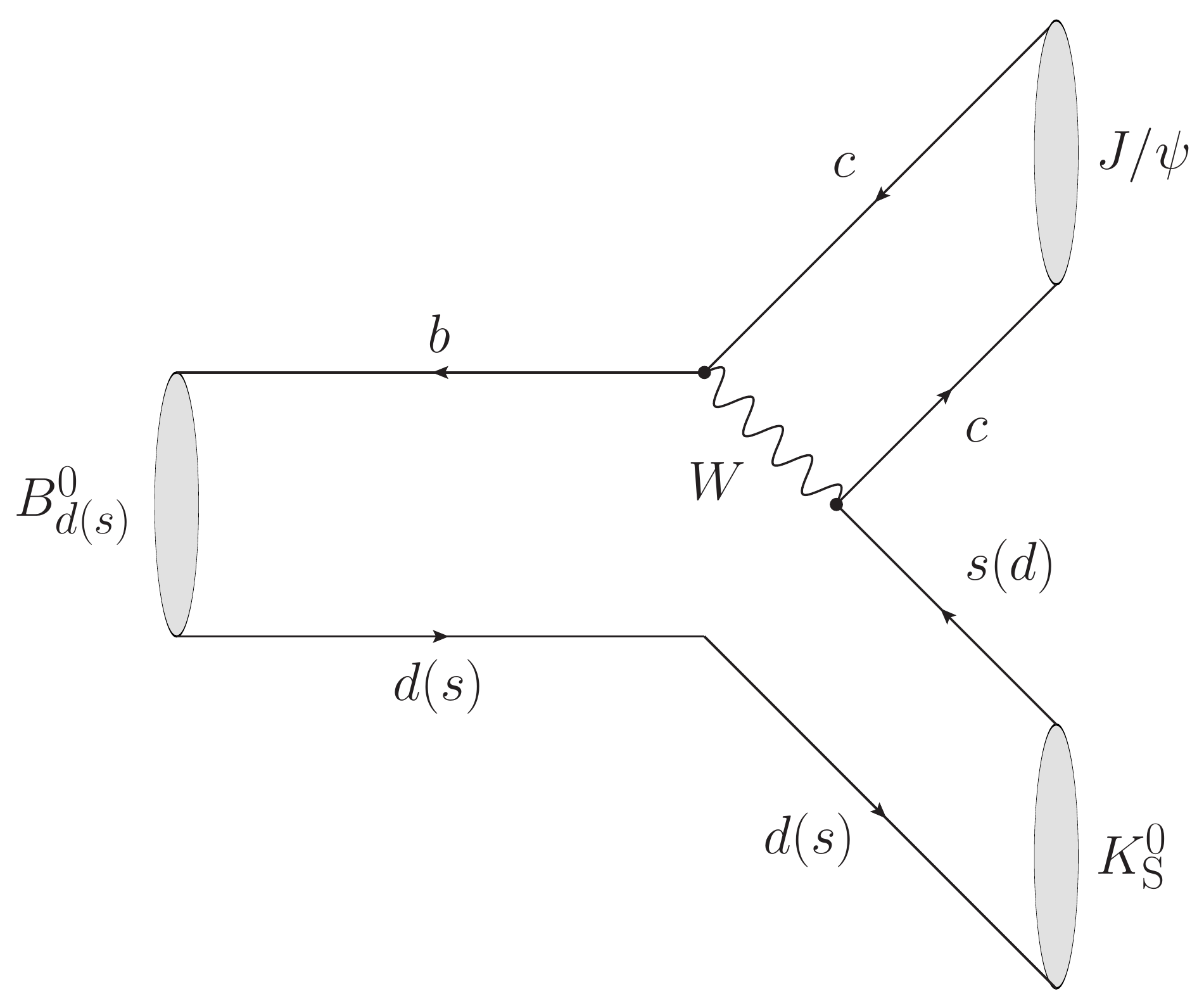}
\hfill
\includegraphics[width=0.49\textwidth]{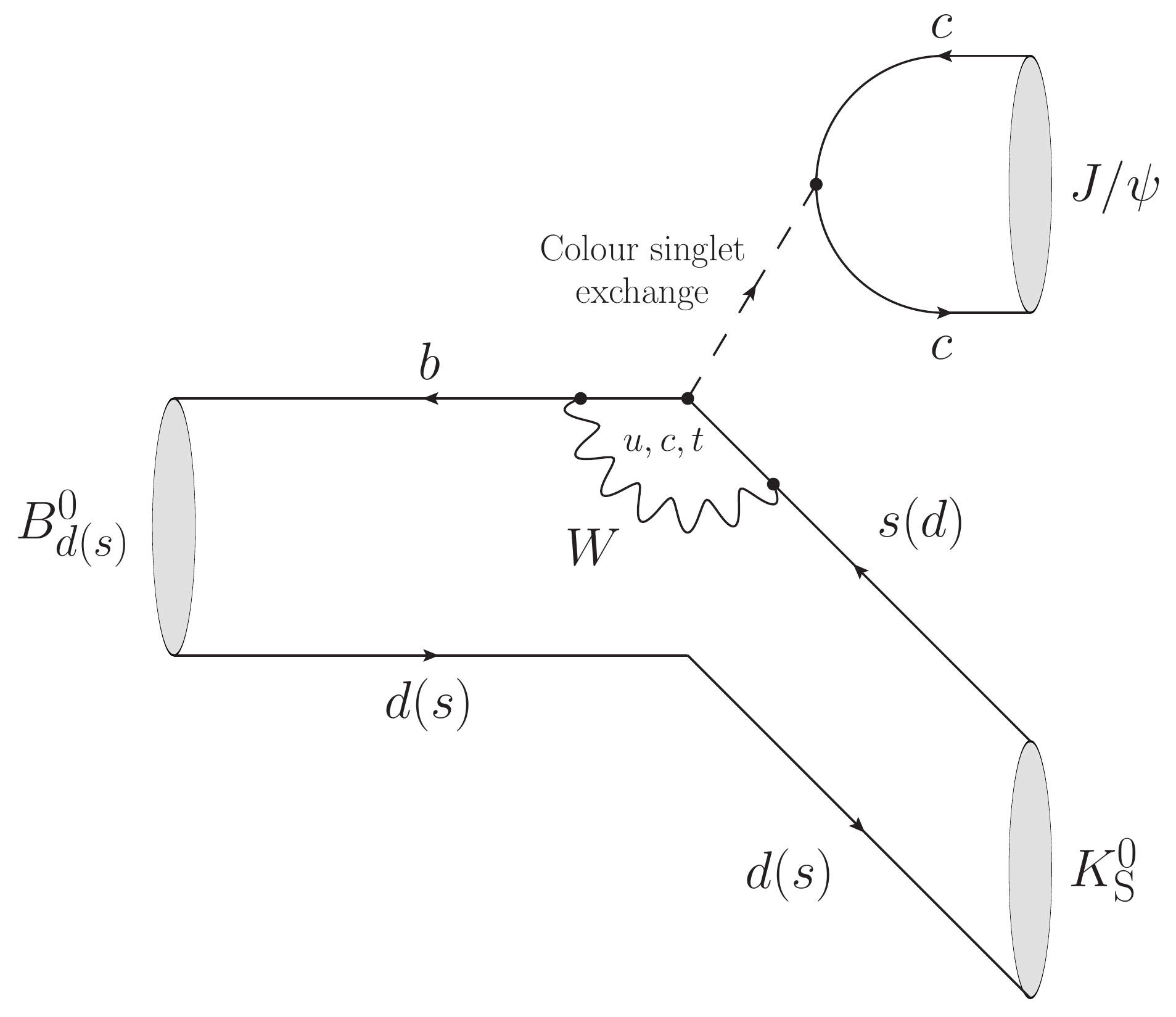}
\caption{Decay topologies contributing to the $B_{d(s)}$\to\jpsi{}\KS channel:
(left) tree diagram and (right) penguin diagram.}
\label{Fig:Feynman}
\end{figure}

From a theoretical perspective, the \Bs\to\jpsi{}\KS mode is the most promising candidate for this task.
It is related to the \Bd\to\jpsi{}\KS mode through the interchange of all $d$ and $s$ quarks
($U$-spin symmetry, a subgroup of $SU(3)$)~\cite{Fleischer:1999nz},
leading to a one-to-one correspondence between all decay topologies of these two modes,
as illustrated in Fig.~\ref{Fig:Feynman}.
Moreover, the \Bs\to\jpsi{}\KS penguin topologies are not CKM suppressed relative to the tree diagram,
as is the case for their \Bd counterparts.
A further discussion regarding the theory of this decay and its 
potential use in LHCb is given in Ref.~\cite{DeBruyn:2010hh, *DeBruyn:2010ge}.

% =====
% What to Measure
To determine the parameters related to the penguin contributions in these decays,
a time-dependent \CP{} violation study of the \Bs\to\jpsi{}\KS mode is required.
The determination of its branching fraction, previously measured by CDF \cite{Aaltonen:2011sy} and LHCb
\cite{LHCb-PAPER-2011-041},
was an important first step, allowing a test of the $U$-spin symmetry assumption that lies 
at the basis of the proposed approach.
The second step towards the time-dependent \CP{} violation study is the measurement
of the effective \Bs\to\jpsi{}\KS lifetime, formally defined as \cite{Fleischer:2010ib}
\begin{equation}\label{Eq:EffLifetimeDef}
\tau_{\jpsi{}\KS}^{\text{eff}} \equiv \frac{\int_0^{\infty} t\:\langle\Gamma(B_s(t)\rightarrow \jpsi{}\KS)\rangle\:\mathrm{d}t}{\int_0^{\infty} \langle\Gamma(B_s(t)\to\jpsi{}\KS)\rangle\:\mathrm{d}t}\:,
\end{equation}
where
\begin{eqnarray}
\langle\Gamma(B_s(t)\to\jpsi{}\KS)\rangle & = & \Gamma(\Bs(t)\to\jpsi{}\KS) + \Gamma(\Bsb(t)\to\jpsi{}\KS)\\
& = & R_{\mathrm H} e^{-\Gamma_{\mathrm H} t} + R_{\mathrm L} e^{-\Gamma_{\mathrm L} t}
\end{eqnarray}
is the untagged decay time distribution,
under the assumption that \CP{} violation in \Bs--\Bsb mixing can be neglected \cite{HFAG}.
Due to the non-zero decay width difference
\mbox{$\Delta\Gamma_s \equiv \Gamma_{\mathrm H} - \Gamma_{\mathrm L} = \DeltaGamma$
\cite{LHCb-PAPER-2013-002}} between the heavy and light \Bs mass eigenstates,
the effective lifetime does not coincide with the \Bs lifetime
\mbox{$\tau_{\Bs} \equiv 1/\Gamma_s = \tauBs$ \cite{LHCb-PAPER-2013-002}},
where $\Gamma_s = (\Gamma_{\mathrm H} + \Gamma_{\mathrm L})/2$ is the average \Bs decay width.
Instead, it depends on the decay mode specific relative contributions
$R_{\mathrm H}$ and $R_{\mathrm L}$.
These two parameters also define the \CP{} observable
\begin{equation}
\mathcal{A}_{\Delta\Gamma_s}\equiv \frac{R_{\mathrm H} - R_{\mathrm L}}{R_{\mathrm H} + R_{\mathrm L}}\:,
\end{equation}
which allows the effective lifetime to be expressed as \cite{Fleischer:2010ib}
\begin{equation}
\tau_{\jpsi{}\KS}^{\text{eff}} = \frac{\tau_{\Bs}}{1-y_s^2}\frac{1+2\: \mathcal{A}_{\Delta\Gamma_s} y_s + y_s^2}{1+\mathcal{A}_{\Delta\Gamma_s} y_s}\:,
\end{equation}
where $y_s \equiv \Delta\Gamma_s/2 \Gamma_s$ is the normalised decay width difference.
For the \Bs\to\jpsi{}\KS mode, the value of $\mathcal{A}_{\Delta\Gamma_s}$ depends on the penguin contributions,
and in particular on their relative weak phase $\phi_s$ \cite{Fleischer:1999nz}.
Using the latest estimates on the size of the \Bs\to\jpsi{}\KS penguin contributions \cite{Fleischer:2012dy} gives
$\mathcal{A}_{\Delta\Gamma_s} = \ADGSM \pm \ADGSME$ and the SM prediction
\begin{equation}\label{Eq:TauEff_SM}
\tau_{\jpsi{}\KS}^{\text{eff}}\Big|_{\text{SM}} = \tauEffSM\:.
\end{equation}
Effective lifetime measurements have been performed for the \mbox{\Bs\to\Kp{}\Km}
\cite{LHCb-PAPER-2012-013} and \mbox{\Bs\to\jpsi{}$f_0(980)$} \cite{LHCb-PAPER-2012-017} decay modes.

%%%%%%%%%%%%%%%%%%%%%%%%%%%%%%%%%%%%%%%%%%%%%%%%%%%%%%%%%%%%%%%%%%%%%%%%%%%
%%%%%%%%%%%%%%%%%%%%%%%%%%%%%%%%%%%%%%%%%%%%%%%%%%%%%%%%%%%%%%%%%%%%%%%%%%%

% =====
% About LHCb
This paper presents the first measurement of the effective \Bs\to\jpsi{}\KS lifetime,
as well as an update of the time-integrated branching fraction measurement in Ref.~\cite{LHCb-PAPER-2011-041},
performed with a data sample, corresponding to an integrated luminosity of $1.0\:\invfb$ of $pp$ collisions,
recorded at a centre-of-mass energy of $7\:\tev$ by the LHCb experiment in 2011.

The \lhcb detector \cite{Alves:2008zz} is a single-arm forward spectrometer covering the pseudorapidity
range $2<\eta <5$, designed for the study of particles containing \bquark or \cquark quarks.
The detector includes a high precision tracking system consisting of a silicon-strip vertex detector surrounding
the $pp$ interaction region, a large-area silicon-strip detector located upstream of a dipole magnet
with a bending power of about $4{\rm\,Tm}$, and three stations of silicon-strip detectors
and straw drift tubes placed downstream.
The combined tracking system has momentum resolution $\Delta p/p$ that varies from 0.4\% at 5\gevc to
0.6\% at 100\gevc, and impact parameter resolution of 20\mum for tracks with high transverse momentum (\pt)
with respect to the beam direction.
Charged hadrons are identified using two ring-imaging Cherenkov detectors \cite{LHCb-DP-2012-003}.
Photon, electron and hadron candidates are identified by a calorimeter system consisting of
scintillating-pad and preshower detectors, an electromagnetic calorimeter and a hadronic calorimeter.
Muons are identified by a system composed of alternating layers of iron and multiwire proportional chambers.

Events are selected by a trigger system \cite{LHCb-DP-2012-004} consisting of a hardware trigger,
which requires muon or hadron candidates with high \pt, followed by a two-stage software trigger. 
In the first stage a partial event reconstruction is performed.
For this analysis, events are required to have either two oppositely charged muons with combined mass
above $2.7\:\gevcc$, or at least one muon or one high-\pt track ($\pt>1.8\:\gevc$) with a
large impact parameter with respect to all $pp$ interaction vertices (PVs).
In the second stage a full event reconstruction is performed and only events 
containing \mbox{\jpsi\to\mup{}\mun} candidates are retained.

The signal simulation samples used for this analysis are generated using
\pythia~6.4~\cite{Sjostrand:2006za} with a specific \lhcb configuration \cite{LHCb-PROC-2010-056}. 
Decays of hadronic particles are described by \evtgen \cite{Lange:2001uf} in which final state
radiation is generated using \photos \cite{Golonka:2005pn}.
The interaction of the generated particles with the detector and its response are implemented using the \geant
toolkit \cite{Allison:2006ve, *Agostinelli:2002hh} as described in Ref.~\cite{LHCb-PROC-2011-006}.

% Data sample and initial selection
%%%%%%%%%%%%%%%%%%%%%%%%%%%%%%%%%%%%%%%%%%%%%%%%%%%%%%%%%%%%%%%%%%%%%%%%%%%
\section{Data samples and initial selection}\label{Sec:Data}
%%%%%%%%%%%%%%%%%%%%%%%%%%%%%%%%%%%%%%%%%%%%%%%%%%%%%%%%%%%%%%%%%%%%%%%%%%%

% =====
% Trigger
Candidate \B\to\jpsi{}\KS decays are reconstructed in the \jpsi\to\mumu and \mbox{\KS\to\pipi} final state.
% =====
% Reconstruction & Loose - Jpsi
Candidate \jpsi\to\mumu decays are required to form a good quality vertex and have a mass in the range
$[3030,3150]\:\mevcc$.
This interval corresponds to about eight times the $\mumu$ mass resolution at the \jpsi mass
and covers part of the $\jpsi$ radiative tail. 
The selected \jpsi candidate is required to satisfy the trigger decision at both software trigger stages.
% =====
% Reconstruction & Loose - KS
The \KS selection requires two oppositely charged particles reconstructed in the tracking stations placed on either side
of the magnet, both with hits in the vertex detector (`long \KS' candidate) or without (`downstream \KS' candidate).
The long (downstream) \KS\to\pipi candidates are required to form a good quality vertex
and have a mass within $35\:(64)\:\mevcc$ of the known $\KS$ mass \cite{PDG2012}.
Moreover, to remove contamination from \L\to\proton{}\pim decays, the reconstructed \proton{}\pim mass
of the long (downstream) \KS candidates is required to be more than $6\:(10)\:\mevcc$ away
from the known \L mass \cite{PDG2012}.
Furthermore, the \KS candidates are required to have a flight distance that is at least five times larger than
its uncertainty.

% =====
% Reconstruction & Loose - Bd
Candidate \B mesons are selected from combinations of \jpsi{} and \KS candidates
with mass $m_{\jpsi\KS}$ in the range $[5180,5520]\:\mevcc$.
The reconstructed mass and decay time are obtained from a kinematic fit \cite{Hulsbergen:2005pu} that
constrains the masses of the \mumu and \pipi pairs to the known \jpsi and \KS masses \cite{PDG2012}, respectively,
and constrains the \B candidate to originate from the PV.
In case the event has multiple PVs, all combinations are considered.
The $\chi^2$ of the fit, which has eight degrees of freedom, is required to be less than $72$ and
the estimated uncertainty on the \B mass must not exceed $30\:\mevcc$.
Candidates are required to have a decay time larger than $0.2\:\ps$.
% =====
% Veto B2JpsiKstar
To remove misreconstructed \mbox{\Bd\to\jpsi{}\Kstarz} background that survives the requirement on
the \KS flight distance, the mass of the long \Bd\to\jpsi{}\KS candidates computed under the \jpsi{}\Kpm{}\pimp
mass hypotheses must not be within $20\:\mevcc$ of the known \Bd mass \cite{PDG2012}.

% Multi-variate selection
%%%%%%%%%%%%%%%%%%%%%%%%%%%%%%%%%%%%%%%%%%%%%%%%%%%%%%%%%%%%%%%%%%%%%%%%%%%
\section{Multivariate selection}\label{Sec:NN_description}
%%%%%%%%%%%%%%%%%%%%%%%%%%%%%%%%%%%%%%%%%%%%%%%%%%%%%%%%%%%%%%%%%%%%%%%%%%%

% =====
% Neural Net
The loose selection described above does not suppress the combinatorial background sufficiently to isolate the
small \Bs\to\jpsi{}\KS signal.
The initial selection is therefore followed by a multivariate analysis,
based on a neural network (NN) \cite{0402093v1}.
The NN classifier's output is used as the final selection variable.

% =====
% sWeights
The NN is trained entirely on data, using the \Bd\to\jpsi{}\KS signal as a proxy for the \mbox{\Bs\to\jpsi{}\KS} decay.
The training sample is taken from the mass windows $[5180,5340]\:\mevcc$ and $[5390,5520]\:\mevcc$,
thus avoiding the \Bs signal region.
A normalisation sample consisting of one quarter of the candidates, selected at random,
is left out of the NN training to allow an unbiased measurement of the \Bd yield.
The signal and background weights are determined using the \sPlot
technique \cite{Pivk:2004ty} and obtained by performing an unbinned maximum likelihood fit
to the mass distribution of the candidates surviving the loose selection criteria.
% =====
% PDF
The fitted probability density function (PDF) is defined as the sum of a \Bd signal component
and a combinatorial background.
The parametrisation of the individual components is described in more detail in the next section.

% =====
% NeuroBayes - Training
Due to the differences in the distributions of the input variables of the NN,
as well as the different initial signal to background ratio,
the multivariate selection is performed separately for the \B candidate samples
containing long and downstream \KS candidates.
In the remainder of this paper, these two datasets will be referred to as the long and downstream \KS sample,
respectively.
The NN classifiers use information about the candidate kinematics, vertex and track quality, 
impact parameter, particle identification information from the RICH and muon detectors, 
as well as global event properties like track and interaction vertex multiplicities.
The variables that are used in the NN are chosen to avoid correlations with the reconstructed \B mass.

% =====
% Optimisation
Final selection requirements on the NN classifier outputs are chosen
to optimise the expected sensitivity to the \Bs signal observation.
The expected signal and background yields entering the calculation of the figure of merit
\cite{Punzi:2003bu} are obtained from the normalisation sample
by scaling the number of fitted \Bd candidates, and by counting the number of events in the mass ranges
$[5180,5240]\:\mevcc$ and $[5400,5520]\:\mevcc$, respectively.
After applying the final requirement on the NN classifier output associated with the long (downstream) \KS sample,
the multivariate selection rejects, relative to the initial selection, 98.7\% (99.6\%) of the background
while keeping 71.5\% (50.2\%) of the \Bd signal.
Due to the worse initial signal to background ratio, the final requirement on the NN classifier output
is much tighter in the downstream \KS sample than in the long \KS sample.

% =====
% Multiple Candidates/PVs
After applying the full selection,
the \B candidate can still be associated with more than one PV in about 1\% of the events.
Likewise, about $0.1\%$ of the selected events have several candidates sharing one or more tracks.
In these cases, respectively one of the surviving PVs and one of the candidates is used at random.

% Measurement of the event yield
%%%%%%%%%%%%%%%%%%%%%%%%%%%%%%%%%%%%%%%%%%%%%%%%%%%%%%%%%%%%%%%%%%%%%%%%%%%
\section{Event yields}\label{Sec:Yields}
%%%%%%%%%%%%%%%%%%%%%%%%%%%%%%%%%%%%%%%%%%%%%%%%%%%%%%%%%%%%%%%%%%%%%%%%%%%

%%%%%%%%%%%%%%%%%%%%%%%%%%%%%%%%%%%%%%%%%%%%%%%%%%%%%%%%%%%%%%%%%%%%%%%%%%%
\begin{figure}[tp]
\center
\begin{minipage}{0.49\textwidth}
\begin{center}
Long \KS
\includegraphics[width=\textwidth]{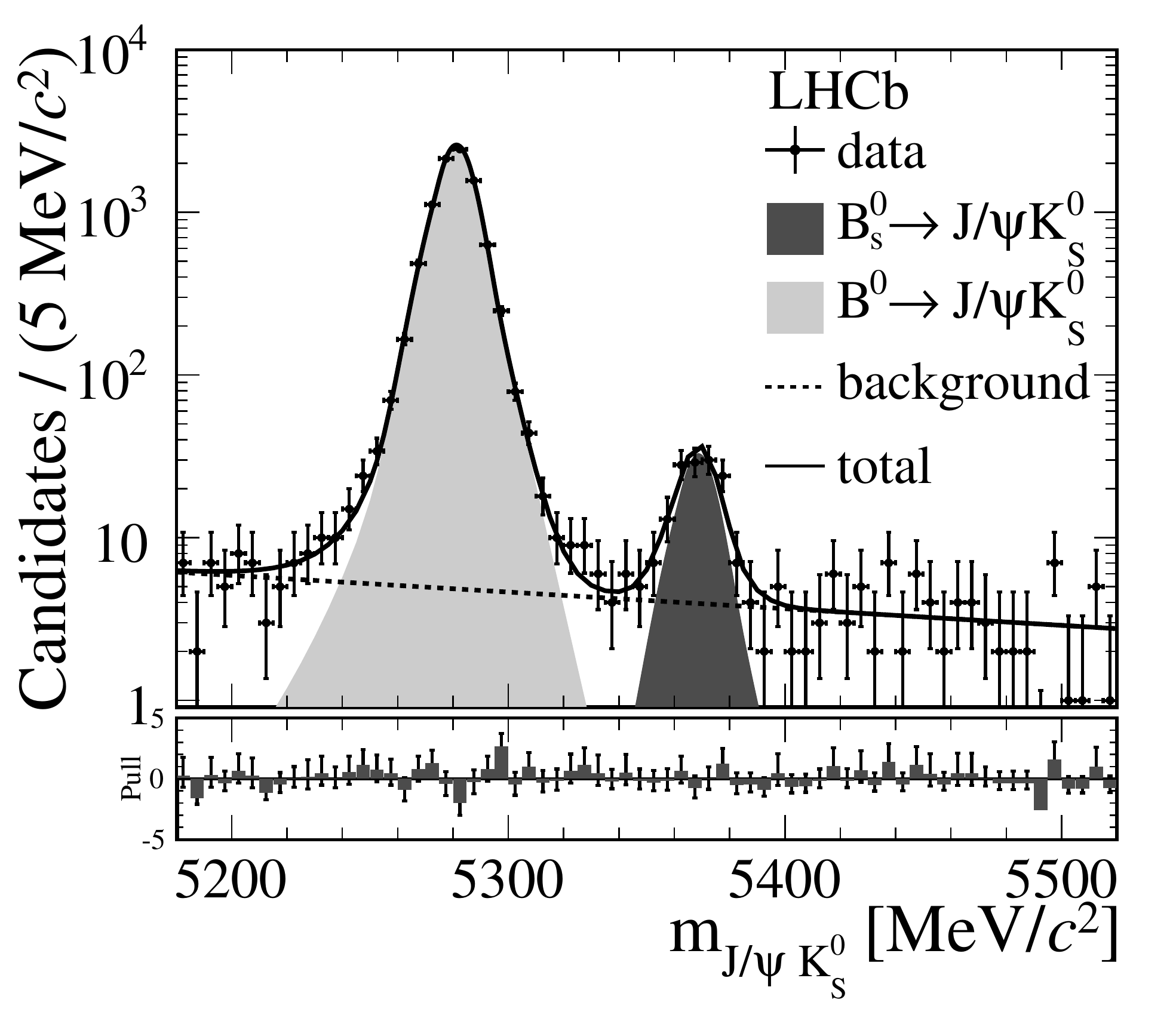}
\end{center}
\end{minipage}
\begin{minipage}{0.49\textwidth}
\begin{center}
Downstream \KS
\includegraphics[width=\textwidth]{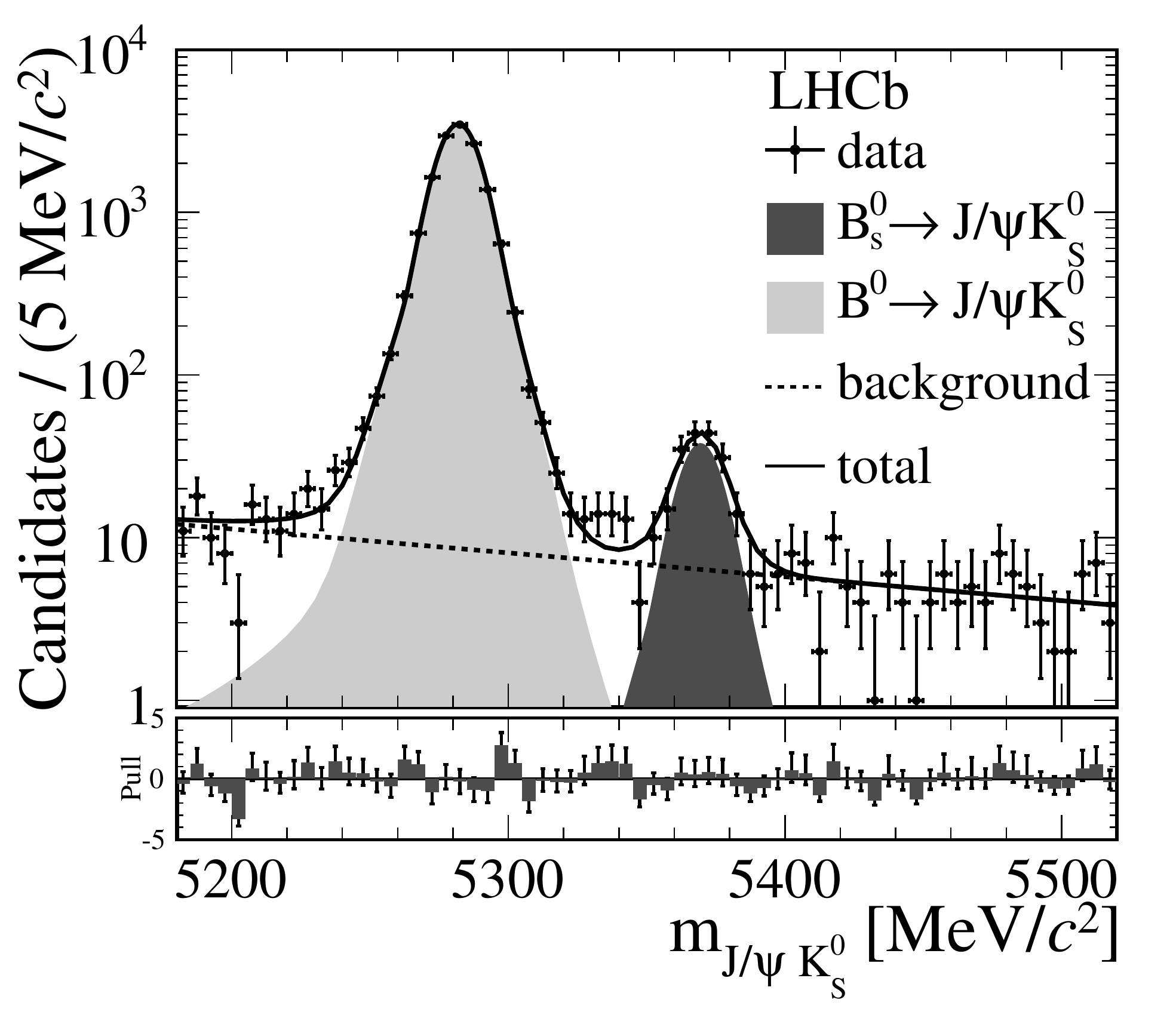}
\end{center}
\end{minipage}
\caption{Fitted \B\to\jpsi{}\KS candidate mass distributions and their associated residual uncertainties (pulls)
for the (left) long and (right) downstream \KS samples,
after applying the final requirement on the NN classifier outputs.
}
\label{Fig:Yield_Results}
\end{figure}
%%%%%%%%%%%%%%%%%%%%%%%%%%%%%%%%%%%%%%%%%%%%%%%%%%%%%%%%%%%%%%%%%%%%%%%%%%%

% =====
% PDF
For the candidates passing the NN requirements, the ratio of \Bs and \Bd yields is determined
from an unbinned maximum likelihood fit to the mass distribution of the reconstructed \B candidates.
The fitted PDF is defined as the sum of a \Bd signal component, a \Bs signal component
and a combinatorial background.
The \Bs component is constrained to have the same shape as the \Bd PDF,
shifted by the known \Bs--\Bd mass difference \cite{LHCb-PAPER-2011-035}.
The mass lineshapes of the \B\to\jpsi{}\KS modes in both data and simulation exhibit non-Gaussian tails
on both sides of their signal peaks due to final state radiation,
the detector resolution and its dependence on the decay angles.
Each individual signal shape is parametrised by a double-sided Crystal Ball (CB) function \cite{Skwarnicki:1986xj}. 
The parameters describing the CB tails are taken from simulation; all other parameters are allowed to vary in the fit.
The background contribution is described by an exponential function.

The results of the fits are shown in Fig.~\ref{Fig:Yield_Results},
and the fitted yields are listed in Table \ref{Tab:Yield_Results}.
The \Bd yield is determined in the normalisation sample and scaled to the full sample, whereas
the \Bs yield is obtained directly from the full sample.
The scaled \Bd yield, obtained from the unbiased sample, differs from the corresponding fit result in the full sample
by $-211 \pm 211$ events for the long \KS sample and by $213 \pm 273$ events for the downstream \KS sample.
Both results are in good agreement, showing that the NN is not overtrained.
The yield ratios obtained from the long and downstream \KS samples are compatible with each other
and are combined using a weighted average.

%%%%%%%%%%%%%%%%%%%%%%%%%%%%%%%%%%%%%%%%%%%%%%%%%%%%%%%%%%%%%%%%%%%%%%%%%%%
\begin{table}[tp]
\center
\caption{Signal yields from the unbinned maximum likelihood fits to the \B\to\jpsi{}\KS candidate mass distributions.
The uncertainties are statistical only.
The yield ratio is calculated from the quantities highlighted in boldface,
where the fitted \Bd yield is first multiplied by a factor of four. 
}
\label{Tab:Yield_Results}
\begin{tabular}{llP{5}P{5}}
Sample & Yield & \multicolumn{1}{c}{Long \KS} & \multicolumn{1}{c}{Downstream \KS}\\
\midrule
\multirow{3}{*}{Normalisation}
& \boldmath\Bd\to\jpsi{}\KS\unboldmath
& \textbf{2205} ! \textbf{47} & \textbf{3651} ! \textbf{61}\\
& \Bs\to\jpsi{}\KS & 21 ! 5 & 49 ! 8\\
& Background & 56 ! 11 & 110 ! 16\\
\midrule
\multirow{3}{*}{Full}
& \Bd\to\jpsi{}\KS & 9031 ! 96 & \text{14,391} ! 122\\
& \boldmath\Bs\to\jpsi{}\KS\unboldmath & \textbf{115} ! \textbf{12} & \textbf{158} ! \textbf{15}\\
& Background & 287 ! 23 & 490 ! 32\\
\midrule
\multicolumn{2}{l}{Yield ratio $R\equiv N_{\Bs\to\jpsi{}\KS}^{\text{Full}}/4N_{\Bd\to\jpsi{}\KS}^{\text{Norm}}$}
& 0.0131 ! 0.0014 & 0.0108 ! 0.0010 \\
\midrule
\multicolumn{2}{l}{Average yield ratio $R$} & \multicolumn{2}{c}{$\YRatio \pm \YRatioE$}\\
\end{tabular}
\end{table}
%%%%%%%%%%%%%%%%%%%%%%%%%%%%%%%%%%%%%%%%%%%%%%%%%%%%%%%%%%%%%%%%%%%%%%%%%%%

% Description of the decay time distribution
%%%%%%%%%%%%%%%%%%%%%%%%%%%%%%%%%%%%%%%%%%%%%%%%%%%%%%%%%%%%%%%%%%%%%%%%%%%
\section{Decay time distribution}\label{Sec:Decay_Time}
%%%%%%%%%%%%%%%%%%%%%%%%%%%%%%%%%%%%%%%%%%%%%%%%%%%%%%%%%%%%%%%%%%%%%%%%%%%

% =====
% Experimental Setup
Following the procedure explained in Ref.~\cite{DeBruyn:2012wj}, 
the effective \Bs\to\jpsi{}\KS lifetime is determined by fitting a single exponential function
$g(t)\propto \exp(-t/\tau_{\text{single}})$
to the decay time distribution of the \mbox{\Bs\to\jpsi{}\KS} signal candidates.
In this analysis, the exponential shape parameter $\tau_{\text{single}}$ is
determined from a two-dimensional unbinned maximum
likelihood fit to the mass and decay time distribution of the reconstructed \B candidates.
The fitted PDF is again defined as the sum of a \Bd signal component, a \Bs signal component
and a combinatorial background.
The freely varying parameters in the fit are the signal and background yields,
and the parameters describing the acceptance, mass and background decay time distributions.

% =====
% Acceptance
The decay time distribution of each of the two signal components needs to be corrected with
a decay time resolution and acceptance model to account for detector effects.
The shape of the acceptance function affecting the \Bs\to\jpsi{}\KS mode is,
like the lineshape of its mass distribution, assumed to be identical to that of the \mbox{\Bd\to\jpsi{}\KS} component.
The acceptance function is obtained directly from the data using the \Bd\to\jpsi{}\KS mode.
Contrary to the \Bs system, the \Bd system has a negligible decay width difference $\Delta\Gamma_d$ 
\cite{PDG2012}.
The decay time distribution of the \Bd\to\jpsi{}\KS channel is therefore fully described by a single exponential
function with known lifetime $\tau_{\Bd} = 1.519\:\text{ps}$ \cite{HFAG}.
Hence, fixing the \Bd lifetime to its known value allows the acceptance parameters to be determined from the fit.

From simulation studies it is found that the decay time acceptance of both signal components
is well modelled by the function
\begin{equation}\label{Eq:Acceptance}
f_{\text{Acc}}\:(t) = \frac{1+\beta\:t}{1+(\lambda\:t)^{-\kappa}}\:.
\end{equation}
The parameter $\beta$ describes the fall in the acceptance at large decay times \cite{LHCb-PAPER-2013-002}.
The parameters $\kappa$ and $\lambda$ model the turn-on curve, caused by the use of decay time biasing triggers,
the initial selection requirements and, most importantly, the NN classifier outputs.

% =====
% Resolution
The decay time resolution for the signal and background components is determined from candidates that have
an unphysical, negative decay time.
Due to the requirement of $0.2\:\text{\ps}$ on the decay time of the \B candidates applied in the initial selection,
such events are not present in the analysed data sample.
Instead, a second sample, that is prescaled and does not have the decay time requirement, is used.
This sample consists primarily of \jpsi mesons produced at the PV which are combined with a random \KS candidate.
The decay time distribution for these events is a good measure of the decay time resolution and
is modelled by the sum of three Gaussian functions sharing a common mean.
Two of the Gaussian functions parametrise the inner core of the resolution function,
while the third describes the small fraction of outliers.

%%%%%%%%%%%%%%%%%%%%%%%%%%%%%%%%%%%%%%%%%%%%%%%%%%%%%%%%%%%%%%%%%%%%%%%%%%%
\begin{figure}[tp]
\begin{minipage}{0.49\textwidth}
\begin{center}
Long \KS
\includegraphics[width=\textwidth]{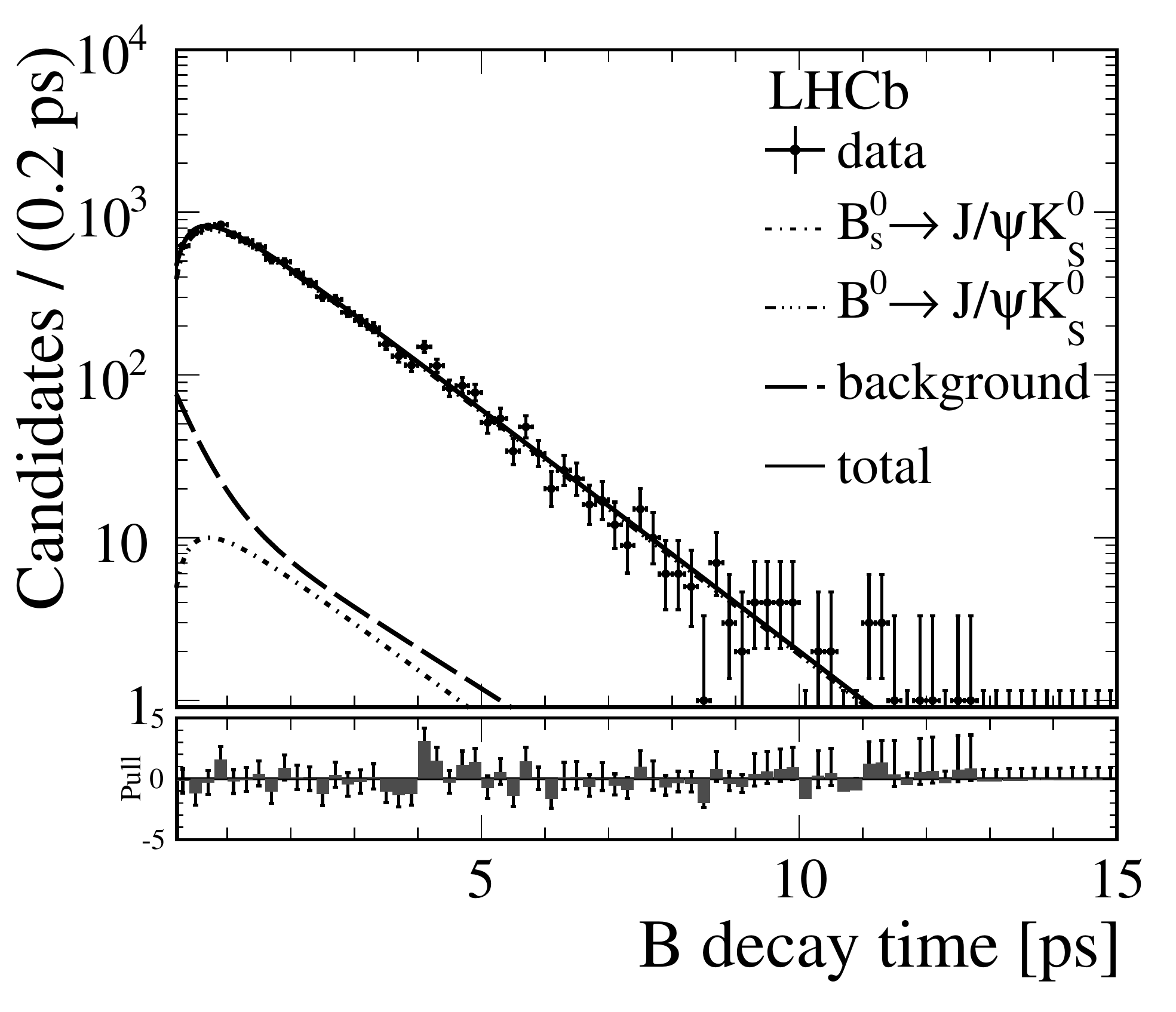}
\end{center}
\end{minipage}
\begin{minipage}{0.49\textwidth}
\begin{center}
Downstream \KS
\includegraphics[width=\textwidth]{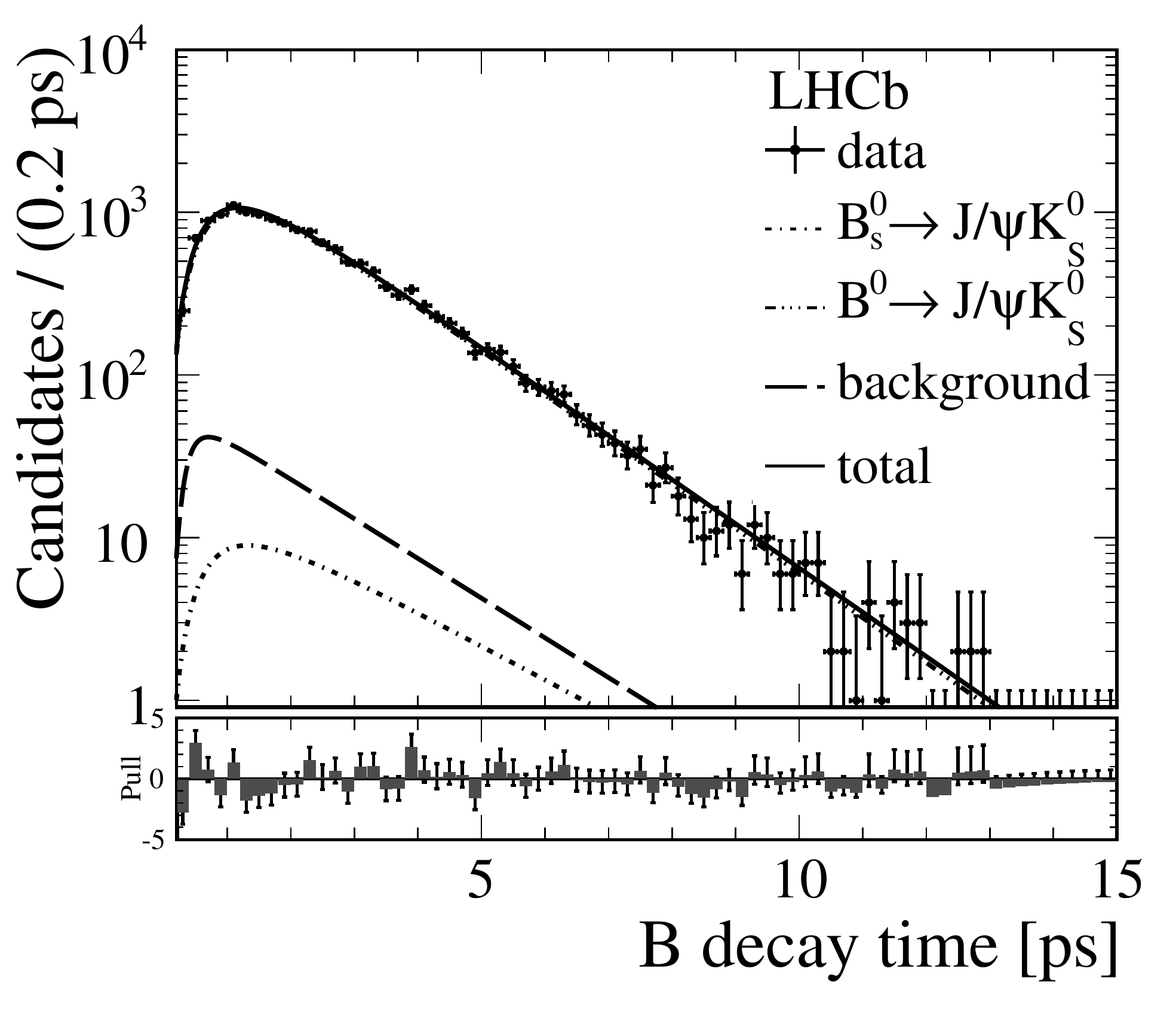}
\end{center}
\end{minipage}
\caption{Fitted \B\to\jpsi{}\KS candidate decay time distributions and their associated residual uncertainties (pulls)
for the (left) long and (right) downstream \KS samples,
after applying the final requirement on the NN classifier outputs.}
\label{Fig:Lifetime_Fit}
\end{figure}
%%%%%%%%%%%%%%%%%%%%%%%%%%%%%%%%%%%%%%%%%%%%%%%%%%%%%%%%%%%%%%%%%%%%%%%%%%%
\begin{figure}[tp]
\begin{minipage}{0.49\textwidth}
\begin{center}
Long \KS
\includegraphics[width=\textwidth]{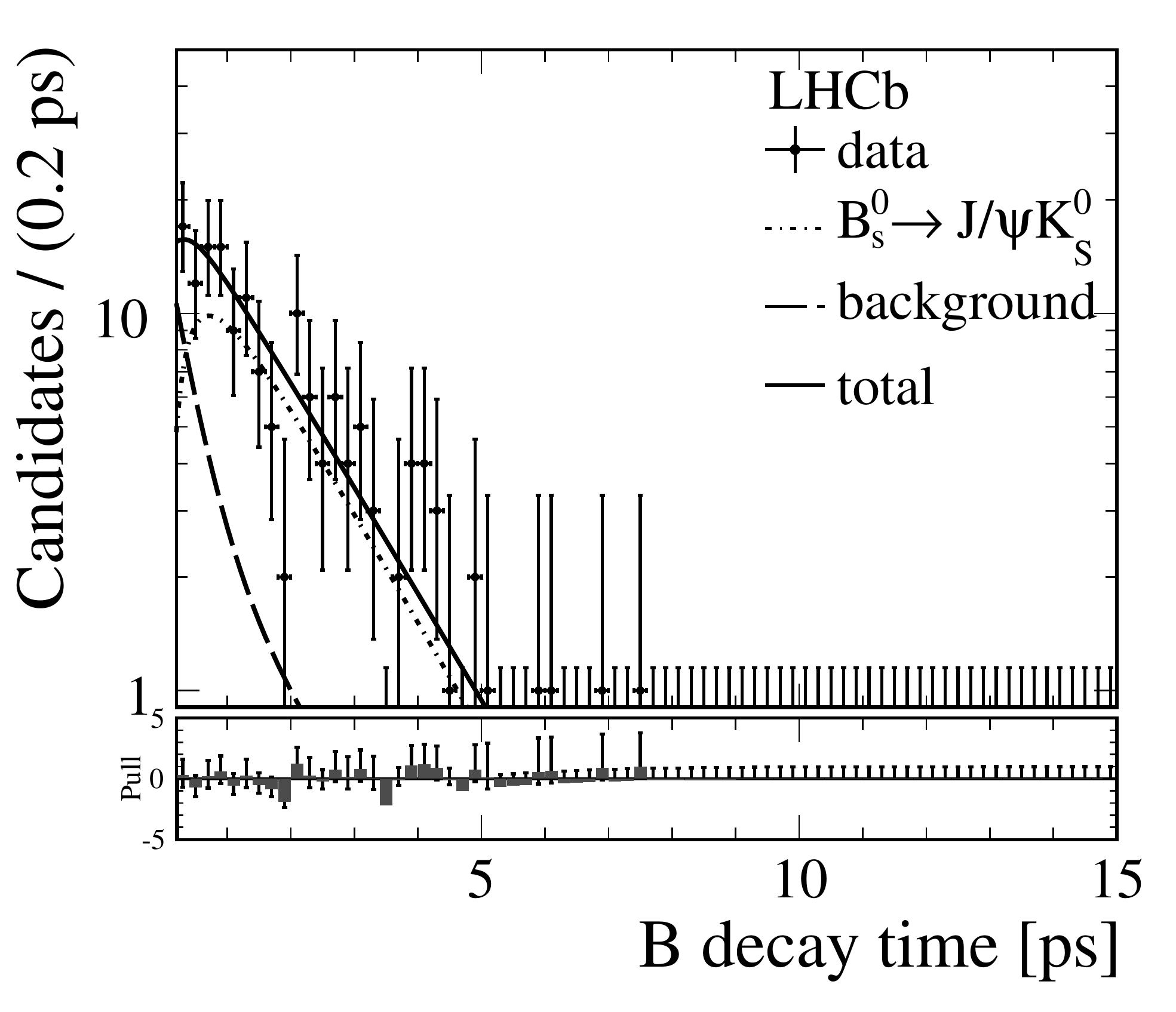}
\end{center}
\end{minipage}
\begin{minipage}{0.49\textwidth}
\begin{center}
Downstream \KS
\includegraphics[width=\textwidth]{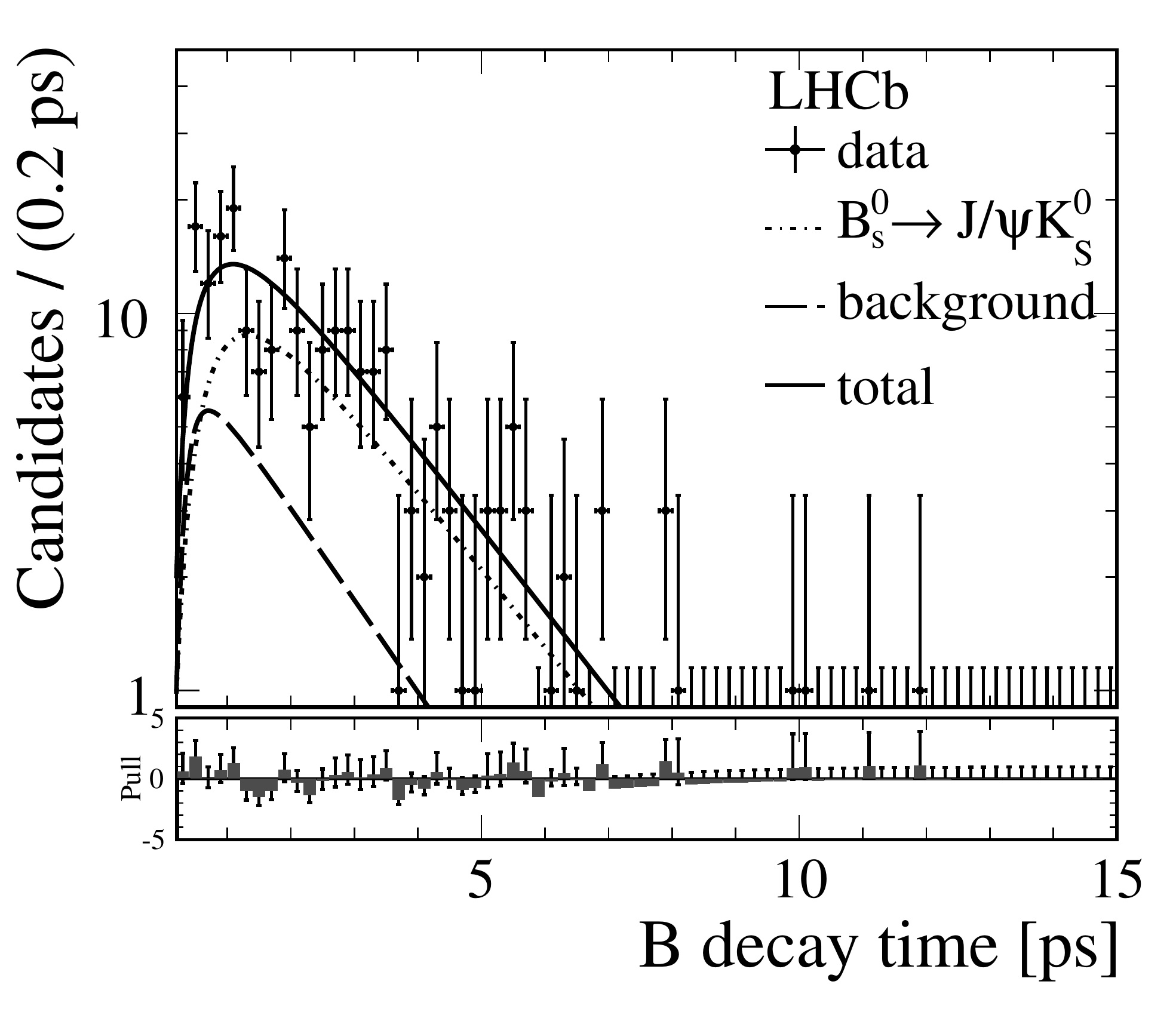}
\end{center}
\end{minipage}
\caption{Fitted \B\to\jpsi{}\KS candidate decay time distributions and their associated residual uncertainties (pulls)
for the (left) long and (right) downstream \KS candidates in the \Bs signal region
$m_{\jpsi{}\KS}\in[5340,5390]\:\mevcc$, after applying the final requirement on the NN classifier outputs.}
\label{Fig:Lifetime_Fit_BsOnly}
\end{figure}
%%%%%%%%%%%%%%%%%%%%%%%%%%%%%%%%%%%%%%%%%%%%%%%%%%%%%%%%%%%%%%%%%%%%%%%%%%%

% =====
% Background
The background decay time distributions are studied directly using the data.
Their shape is obtained from background candidates that are isolated using the background weights determined
by the \sPlot technique, and cross-checked using the high mass sideband.
The exact values of the shape parameters are determined in the nominal fit.
Because of the differences induced by the multivariate selection, the background decay time distribution
of the long and downstream \KS samples cannot be parametrised using the same background model.
For the long \KS sample, the background is modelled by two exponential functions,
describing a short-lived and a long-lived component, respectively.
In the downstream \KS sample such a short-lived component is not present due to the tighter requirement
on the NN classifier output.
Its decay time distribution is better described by a single exponential shape corrected by the acceptance function
in Eq.~\eqref{Eq:Acceptance} with independent parameters $(\kappa', \alpha', \beta')$.
The parameter $\beta'$ is set to zero because we also fit the lifetime of the single exponential function itself,
and the combination of both parameters would result in ambiguous solutions.

% =====
% Final Fit
The decay time distributions resulting from the two-dimensional fits are shown in Figs.~\ref{Fig:Lifetime_Fit} and
\ref{Fig:Lifetime_Fit_BsOnly} for candidates in the full mass range \mbox{$m_{\jpsi{}\KS}\in[5180,5520]\:\mevcc$} and
in the \Bs signal region \mbox{$m_{\jpsi{}\KS}\in[5340,5390]\:\mevcc$}, respectively.
The fitted values are $\tau_{\text{single}} = 1.54 \pm 0.17\:\text{ps}$ and
$\tau_{\text{single}} = 1.96 \pm 0.17\:\text{ps}$ for the long and downstream \KS sample, respectively.
The $1.7\sigma$ difference between both results is understood as a statistical fluctuation.
The two main fit results are therefore combined using a weighted average, leading to
\begin{equation*}
\tau_{\text{single}} = \EffRaw \pm \EffRawE\:\text{ps}\:,
\end{equation*}
where the uncertainty is statistical only.
The event yields obtained from the two-dimensional fits are compatible with the results
quoted in Table \ref{Tab:Yield_Results}.

% Corrections and systematics
%%%%%%%%%%%%%%%%%%%%%%%%%%%%%%%%%%%%%%%%%%%%%%%%%%%%%%%%%%%%%%%%%%%%%%%%%%%
\section{Corrections and systematic uncertainties}\label{Sec:Syst}
%%%%%%%%%%%%%%%%%%%%%%%%%%%%%%%%%%%%%%%%%%%%%%%%%%%%%%%%%%%%%%%%%%%%%%%%%%%

% =====
% Overview
A number of systematic uncertainties affecting the relative branching fraction
\mbox{\BR(\Bs\to\jpsi{}\KS)/\BR(\Bd\to\jpsi{}\KS)} and the effective lifetime are considered.
The sources affecting the ratio of branching fractions are discussed first, followed by those contributing to the
effective lifetime measurement.
\newline

% =====
% Yield Ratio

% =====
% Fit Model & Mass Resolution
The largest systematic uncertainty on the yield ratio comes from the mass shape model,
and in particular from the uncertainty
on the fraction of the \Bd\to\jpsi{}\KS component's high mass tail extending below the \Bs signal.
The magnitude of this effect is studied by allowing both tails of the CB shapes to vary in the fit.
The largest observed deviation in the yield ratios is 3.4\%, which is taken as a systematic uncertainty.
The mass resolution, and hence the widths of the CB shapes, is assumed to be identical for the \Bd and \Bs signal
modes, but could in principle depend on the mass of the reconstructed \B candidate.
This effect is studied by multiplying the CB widths of the \Bs signal PDF by different scale factors,
obtained by comparing \Bd and \Bs signal shapes in simulation.
The largest observed difference in the yield ratios is 1.4\%, which is taken as a systematic uncertainty.
Varying the \Bs--\Bd mass difference within its uncertainty has negligible effect on the yield ratios.

% =====
% Selection Efficiencies
The selection procedure is designed to be independent of the reconstructed \B mass.
Simulated data is used to check this assumption, and to evaluate the difference in selection efficiency
arising from the different shapes of the \Bd\to\jpsi{}\KS and \Bs\to\jpsi{}\KS decay time distributions.
The ratio of total selection efficiencies is equal to \mbox{$\SystSelEffCorr \pm \SystSelEffCorrE$},
and is used to correct the yield ratio.

% =====
% Optimisation
The stability of the multivariate selection is verified by comparing different training schemes and optimisation
procedures, as well as by calculating the yield ratios for different subsets of the long and downstream \KS sample.
All of these tests give results that are compatible with the measured ratio.

% =====
% Summary
The corrections and systematic uncertainties affecting the branching fraction ratio are listed in
Table \ref{Tab:Systematics}.
The total systematic uncertainty is obtained by adding all the uncertainties in quadrature.
\newline

%%%%%%%%%%%%%%%%%%%%%%%%%%%%%%%%%%%%%%%%%%%%%%%%%%%%%%%%%%%%%%%%%%%%%%%%%%%
\begin{table}[tp]
\center
\caption{Corrections and systematic uncertainties on the yield ratio.}
\label{Tab:Systematics}
\begin{tabular}{lP{4}}
Source & \multicolumn{1}{c}{Value}\\
\midrule
Fit model & \SystFitModel ! \SystFitModelE\\
\Bs mass resolution & \SystMassRes ! \SystMassResE\\
Selection efficiency & \SystSelEffCorr ! \SystSelEffCorrE\\
\midrule
Total correction $f_{\text{corr}}^{\BR}$ & \SystTot ! \SystTotE\\
\end{tabular}
\end{table}
%%%%%%%%%%%%%%%%%%%%%%%%%%%%%%%%%%%%%%%%%%%%%%%%%%%%%%%%%%%%%%%%%%%%%%%%%%%

% =====
% Effective Lifetime

% =====
% Modelling
The main systematic uncertainties affecting the effective \Bs\to\jpsi{}\KS lifetime arise from modelling
the different components of the decay time distribution.
Their amplitudes are evaluated by comparing the results from the nominal fit
to similar fits using alternative parametrisations.
All tested fit models give compatible results.
The largest observed deviations in $\tau_{\text{single}}$ are 3.9\% due to modelling of the background decay time
distribution, 0.47\% due to the acceptance function and 0.39\% due to the reconstructed \B mass description,
all of which are assigned as systematic uncertainties.
Variations in the decay time resolution model are found to have negligible impact on $\tau_{\text{single}}$.

% =====
% Bd Lifetime
The assumed value of the \Bd lifetime has a significant impact on the shape of the acceptance function,
and the $\beta$ parameter in particular, which in turn affects the fitted value of $\tau_{\text{single}}$.
This effect is studied by varying the \Bd lifetime within its uncertainty \cite{PDG2012}.
The largest observed deviation in $\tau_{\text{single}}$ is 0.52\%, which is taken as a systematic uncertainty.

% =====
% Fit Method
The fit method is tested on simulated data using large sets of pseudo-experiments,
which have the same mass and decay time distributions as the data.
Different datasets are generated using the fitted two-dimensional signal and background distributions,
and $\tau_{\text{single}}$ is then again fitted to these pseudo-experiments.
The fit result is compared with the input value to search for possible biases.
From the spread in the fitted values and the accompanying residual distributions, a small bias is found.
This bias is attributed to the limited size of the background sample,
and the resulting difficulty to constrain the background decay time parameters.
A correction factor of $\SystFitMethod \pm \SystFitMethodE$ is assigned to account for this potential bias.

% =====
% Definition
Due to the presence of a non-trivial acceptance function, the result of fitting a single exponential to the untagged
\Bs decay time distribution does mathematically not agree with the formal definition of the effective lifetime
in Eq.~\eqref{Eq:EffLifetimeDef}, as explained in Ref.~\cite{DeBruyn:2012wj}.
The size and sign of the difference between $\tau_{\text{single}}$ and $\tau_{\jpsi{}\KS}^{\text{eff}}$ depend on
the values of $\tau_{\Bs}$, $y_s$, $\mathcal{A}_{\Delta\Gamma_s}$, and the shape of the acceptance function.
The difference is calculated with pseudo-experiments that sample the acceptance parameters, $\tau_{\Bs}$
and $y_s$ from Gaussian distributions related to their respective fitted and known values.
Since $\mathcal{A}_{\Delta\Gamma_s}$ is currently not constrained by experiment,
it is sampled uniformly from the interval $[{-1}, 1]$.
The average difference between $\tau_{\jpsi{}\KS}^{\text{eff}}$ and $\tau_{\text{single}}$,
obtained using the acceptance function affecting the long (downstream) \KS sample,
is found to be $-0.001\:\text{ps}$ $(-0.003\:\text{ps})$.
A correction factor of $\SystTauDef \pm \SystTauDefE$ is assigned to account for this bias.

% =====
% Asymmetry
The presence of a production asymmetry between the \Bs and \Bsb mesons could potentially alter the measured
value of the effective lifetime,
but even for large estimates of the size of this asymmetry, the effect is found to be negligible.

% =====
% Momentum & Decay length scale
Finally, the systematic uncertainties in the momentum and the decay length scale propagate to the effective lifetime.
The size of the former contribution is evaluated
by recomputing the decay time while varying the momenta of the final state particles within their uncertainty.
The systematic uncertainty due to the decay length scale mainly comes from the track-based alignment.
Both effects are found to be negligible.

% =====
% Stability
The stability of the fit is verified by comparing the nominal results with those obtained using different fit ranges,
or using only subsets of the long and downstream \KS samples.
All these tests give compatible results.

% =====
% Summary
The corrections and systematic uncertainties affecting the effective \Bs\to\jpsi{}\KS lifetime are listed in
Table \ref{Tab:Systematics_Lifetime}.
The total systematic uncertainty is obtained by adding all the uncertainties in quadrature.

%%%%%%%%%%%%%%%%%%%%%%%%%%%%%%%%%%%%%%%%%%%%%%%%%%%%%%%%%%%%%%%%%%%%%%%%%%%
\begin{table}[tp]
\center
\caption{Corrections and systematic uncertainties on the effective \Bs\to\jpsi{}\KS lifetime.}
\label{Tab:Systematics_Lifetime}
\begin{tabular}{lP{4}}
Source & \multicolumn{1}{c}{Value}\\
\midrule
Background model & \SystBkg ! \SystBkgE\\
Acceptance model & \SystAcc ! \SystAccE\\
Mass model & \SystMass ! \SystMassE\\
\Bd lifetime & \SystBdLife ! \SystBdLifeE\\
Fit method & \SystFitMethod ! \SystFitMethodE\\
Effective lifetime definition & \SystTauDef ! \SystTauDefE\\
\midrule
Total correction $f_{\text{corr}}^{\text{eff}}$ & \SystTauTot ! \SystTauTotE\\
\end{tabular}
\end{table}
%%%%%%%%%%%%%%%%%%%%%%%%%%%%%%%%%%%%%%%%%%%%%%%%%%%%%%%%%%%%%%%%%%%%%%%%%%%

% Conclusion
%%%%%%%%%%%%%%%%%%%%%%%%%%%%%%%%%%%%%%%%%%%%%%%%%%%%%%%%%%%%%%%%%%%%%%%%%%%
\section{Results and conclusion}\label{Sec:Conclusion}
%%%%%%%%%%%%%%%%%%%%%%%%%%%%%%%%%%%%%%%%%%%%%%%%%%%%%%%%%%%%%%%%%%%%%%%%%%%
Using the measured ratio $R=\YRatio \pm \YRatioE$ of \Bs\to\jpsi{}\KS and \Bd\to\jpsi{}\KS yields,
the correction factor $f_{\text{corr}}^{\BR} = \SystTot \pm \SystTotE$,
and the ratio of hadronisation fractions \mbox{$f_s/f_d = \fsfd$}
\cite{LHCb-PAPER-2012-037}, the ratio of branching fractions is computed to be
\begin{eqnarray*}
\frac{\BR(\Bs\to\jpsi{}\KS)}{\BR(\Bd\to\jpsi{}\KS)}
& = & R \times f_{\text{corr}}^{\BR} \times \frac{f_d}{f_s}\\
& = & \BRBsBd \pm \BRBsBdStat\:\text{(stat)} \pm\BRBsBdSyst\:\text{(syst)} \pm\BRBsBdfds\:(f_s/f_d)\:,
\end{eqnarray*}
where the quoted uncertainties are statistical, systematic, and due to the uncertainty in $f_s/f_d$, respectively.

Using the known \Bd\to\jpsi{}\Kz branching fraction  \cite{PDG2012},
the ratio of branching fractions can be converted into
a measurement of the time-integrated \Bs\to\jpsi{}\KS branching fraction.
Taking into account the different rates of \Bu{}\Bub and \Bd{}\Bdb pair production
at the \FourS resonance $\Gamma(\Bu{}\Bub)/\Gamma(\Bd{}\Bdb) = \U4S$ \cite{PDG2012},
the above result is multiplied by the corrected value
\mbox{$\BR(\Bd\to\jpsi{}\Kz) = (\BFBd \pm \BFBdE)\times 10^{-4}$} and gives
\begin{eqnarray*}
&& \BR(\Bs\to\jpsi{}\KS) = \\
&& \left[\BFBs \pm \BFBsStat\:\text{(stat)} \pm \BFBsSyst\:\text{(syst)}
\pm \BFBsfds\:(f_s/f_d) \pm \BFBsPDG\:(\BR(\Bd\to\jpsi{}\Kz))\right]\times10^{-5}\:,
\end{eqnarray*}
where the last uncertainty comes from the \Bd\to\jpsi{}\Kz branching fraction.
This result is compatible with, and more precise than, previous measurements
\cite{Aaltonen:2011sy,LHCb-PAPER-2011-041}, and supersedes the previous \lhcb measurement.
The branching fraction is consistent with expectations from $U$-spin symmetry \cite{LHCb-PAPER-2011-041}.
\newline

Using $\tau_{\text{single}} = \EffRaw \pm \EffRawE\:\text{ps}$
and the correction factor $f_{\text{corr}}^{\text{eff}} = \SystTauTot \pm \SystTauTotE$,
the effective \Bs\to\jpsi{}\KS lifetime is given by
\begin{eqnarray*}
\tau_{\jpsi{}\KS}^{\text{eff}} 
& = & f_{\text{corr}}^{\text{eff}} \times \tau_{\text{single}}\\
& = & \TauBsEff \pm \TauBsEffStat\:(\text{stat}) \pm \TauBsEffSyst\:(\text{syst})\:\text{ps}\:.
\end{eqnarray*}
This is the first measurement of this quantity.
The result is compatible with the SM prediction given in Eq.~\eqref{Eq:TauEff_SM}.

% Acknowledgements
\section*{Acknowledgements}

\noindent We express our gratitude to our colleagues in the CERN
accelerator departments for the excellent performance of the LHC. We
thank the technical and administrative staff at the LHCb
institutes. We acknowledge support from CERN and from the national
agencies: CAPES, CNPq, FAPERJ and FINEP (Brazil); NSFC (China);
CNRS/IN2P3 and Region Auvergne (France); BMBF, DFG, HGF and MPG
(Germany); SFI (Ireland); INFN (Italy); FOM and NWO (The Netherlands);
SCSR (Poland); ANCS/IFA (Romania); MinES, Rosatom, RFBR and NRC
``Kurchatov Institute'' (Russia); MinECo, XuntaGal and GENCAT (Spain);
SNSF and SER (Switzerland); NAS Ukraine (Ukraine); STFC (United
Kingdom); NSF (USA). We also acknowledge the support received from the
ERC under FP7. The Tier1 computing centres are supported by IN2P3
(France), KIT and BMBF (Germany), INFN (Italy), NWO and SURF (The
Netherlands), PIC (Spain), GridPP (United Kingdom). We are thankful
for the computing resources put at our disposal by Yandex LLC
(Russia), as well as to the communities behind the multiple open
source software packages that we depend on.

%\clearpage
\phantomsection
\addcontentsline{toc}{section}{References}
\bibliographystyle{LHCb}
\bibliography{main}

\end{document}